\documentstyle[aps,prl,amsfonts,multicol]{revtex}

\begin{document}
\newcommand{\Z}{{\Bbb Z}}
\newcommand{\w}{{\omega}}
\newcommand{\QFT}{{\cal F}}
\newcommand{\cU}{{\cal U}}
\newcommand{\cI}{{\cal I}}
\newcommand{\cW}{{\cal W}}
\newcommand{\cR}{{\cal R}}

\draft
\title{Quantum Algorithm for Generalized Deutsch-Jozsa Problem}
\author{Dong Pyo Chi,\cite{dpchi}
        Jinsoo Kim,\cite{jskim} and
        Soojoon Lee\cite{level}
}
\address{Department of Mathematics, Seoul National University,
 Seoul 151-742, Korea}
\date{\today}
\maketitle

\begin{abstract}
We generalize the Deutsch-Jozsa problem and
present a quantum algorithm
that can solve the generalized Deutsch-Jozsa problem
by a single evaluation of a given function.
We discuss the initialization of an auxiliary register
and present a generalized Deutsch-Jozsa algorithm
that requires no initialization of an auxiliary register.
\end{abstract}

\pacs{PACS numbers: 03.67.Lx, 03.65.Bz}


The computational power of quantum computer has much been explored
since the early work of Deutsch and Jozsa \cite{DJ}.
They presented a simple promise problem,
which is now called the Deutsch-Jozsa problem, that can be
solved efficiently without error on quantum computer but that
requires exhaustive search to solve deterministically without
error in a classical setting.
The {\em Deutsch-Jozsa problem} is to determine whether
a Boolean function $f:\Z_{2^n} \rightarrow \Z_2$
is nonconstant or non-balanced
where $f$ is said to be {\em balanced}
if $f(x)=0$ for exactly half of the input values.
The Deutsch-Jozsa algorithm \cite{DJ} can solve this problem
by a single evaluation of $f$ on quantum computer.
This algorithm consists of the successive application of
the operators $\cW_n\otimes \cI$, $\cU_f$, and $\cW_n\otimes \cI$
to two quantum registers with their initial state being
$ \left| 0^n \right\rangle \otimes \cW_1 \left| 1 \right\rangle $
where $\cW_j$ is the $j$-qubit Walsh-Hadamard operator and
$\cU_f$ is the function-evaluation operator defined by
$ \left| x \right\rangle \otimes \left| y \right\rangle \mapsto
\left| x \right\rangle \otimes \left| y + f(x) \right\rangle$.
The resulting state becomes
$\frac{1}{N} \sum_{x,y=0}^{N-1} (-1)^{x \cdot y + f(x)}
 \left| y \right\rangle \otimes
 \cW_1 \left| 1 \right\rangle $
where $x \cdot y$ stands for the scalar product modulo $2$ in $\Z_2^n$,
that is, $x \cdot y \equiv \sum_{j=0}^{n-1}x_j y_j \, (\bmod \,2)$
for $x=\sum_{j=0}^{n-1}x_j 2^j$ and $y=\sum_{j=0}^{n-1}y_j 2^j$
with $x_j, y_j \in \Z_2$.
At this stage discarding the second one-qubit register
we perform a measurement on the first $n$-qubit register and conclude
that $f$ is non-balanced
if the outcome is $ \left| 0^n \right\rangle $ and
that $f$ is nonconstant otherwise.

Allowing $f$ to have its values on $\Z_M$
and accordingly modifying the concept of balancedness
we generalize the Deutsch-Jozsa problem.
We say that a function $f:\Z_N \rightarrow \Z_M$ is
{\em evenly distributed}
if $f$ has equally spaced $K$ values
and is a $\nu$-to-one function where $\nu=\frac{N}{K}$.
That is, if $f$ is evenly distributed, then
there exists $t \ge0$ such that the period of the
range of $f$ is $\mu = \frac{M}{K} $ with a possible initial
shift $t$. In other words,
$ \{ f(x) \, :\, x \in \Z_N \} = \{ j \mu +t \, :\, j \in \Z_K \}$
and $| A_0 | = |A_1| = \cdots = |A_{K-1}|$ where $ A_j = \{ x \in
\Z_N \, :\, f(x) = j \mu+t \}$ for $j \in \Z_K$.
The {\em generalized Deutsch-Jozsa problem} is
to determine whether $f$ is
nonconstant or not evenly distributed.
When $f$ is onto, $f$ is an evenly distributed function
if and only if $f$ is a $\nu$-to-one function.
Thus if $M=K$
then the generalized Deutsch-Jozsa problem is equivalent to
determining whether $f$ is nonconstant or non-$\nu$-to-one.
We remark that the $\nu$-to-one function appears
in collision and claw problems
\cite{BHT} under the assumption that $f$ is onto.

When $K$ is known
we need $\nu+1$ evaluations of $f$ classically in worst case
in order to solve the generalized Deutsch-Jozsa problem.
Unless $K$ is known,
any classical algorithm for this problem would require $\frac{N}{2}+1$
evaluations of $f$ in worst case before determining the answer
with certainty.
Thus the generalized Deutsch-Jozsa problem has
the same computational complexity as
that of the original Deutsch-Jozsa problem.


Actually the original Deutsch-Jozsa algorithm can solve
the generalized Deutsch-Jozsa problem
by slightly modifying the initial state of the second register.
For simplicity, we assume that $N$ and $M$ are powers of 2, that
is, $N=2^n$ and $M=2^m$ for some positive integers $n$ and $m$.
We prepare two quantum registers, in which
the first $n$-qubit register called the {\em control register}
is used to store the states we wish to interfere and
the second $m$-qubit register called the {\em auxiliary register}
is used to draw relative phase changes in the first register.
We initialize the control register by $ \left| 0^n \right\rangle $
and the auxiliary register by
$ \left| \Psi \right\rangle  = \QFT \left| -\xi \right\rangle
= \frac{1}{\sqrt{M}} \sum_{z=0}^{M-1} \w_M^{-\xi z}
  \left| z \right\rangle $
for a nonzero $\xi\in\Z_M$
where $\QFT$ is the quantum Fourier transform and
$\w_M=e^{2\pi i/M}$ is a primitive $M$-th root of unity.
We proceed the following algorithm:
(i) Apply $\cW_n \otimes \cI$
(ii) Apply $\cU_f$
(iii) Apply $\cW_n \otimes \cI$.
Then the state evolves as follows:
\begin{eqnarray}
  \left| 0^n \right\rangle \otimes  \left| \Psi \right\rangle
&\stackrel{\cW_n \otimes \cI}{\longrightarrow}&
 \frac{1}{\sqrt{N}} \sum_{x=0}^{N-1}  \left| x \right\rangle
 \otimes  \left| \Psi \right\rangle
 \nonumber \\
&\stackrel{\cU_f}{\longrightarrow}&
 \frac{1}{\sqrt{NM}} \sum_{x=0}^{N-1} \sum_{z=0}^{M-1}
 \w_M^{-\xi z}  \left| x \right\rangle \otimes
 \left| z+f(x) \right\rangle
 \nonumber \\
&&=
  \frac{1}{\sqrt{NM}} \sum_{x=0}^{N-1}
  \sum_{z=0}^{M-1} \w_M^{-\xi z} \w_M^{\xi f(x)}
  \left| x \right\rangle \otimes \left| z \right\rangle
  \nonumber \\
&&=
  \frac{1}{\sqrt{N}} \sum_{x=0}^{N-1} \w_M^{\xi f(x)}
  \left| x \right\rangle
  \otimes  \left| \Psi \right\rangle
  \nonumber \\
&\stackrel{\cW_n\otimes \cI}{\longrightarrow}&
 \sum_{y=0}^{N-1}
 \left( \frac{1}{N}\sum_{x=0}^{N-1}
 (-1)^{x \cdot y} \w_M^{\xi f(x)} \right)  \left| y \right\rangle
 \otimes  \left| \Psi \right\rangle  .
 \label{Alg:GDJ1}
\end{eqnarray}
Let $S_y$ be the inner summation in the final state of (\ref{Alg:GDJ1});
$S_y = \frac{1}{N}\sum_{x=0}^{N-1} (-1)^{x \cdot y} \w_M^{\xi f(x)}$.
If $f$ is constant, then we obtain
\begin{eqnarray*}
S_y&=&\frac{1}{N}\w_M^{\xi f(0)} \sum_{x=0}^{N-1} (-1)^{x\cdot y} \\
   &=&\cases{0 & when $y \ne 0$, \cr  \w_M^{\xi f(0)} & when $y = 0$.}
\end{eqnarray*}
If $f$ is evenly distributed, then for $y=0$ we have
\begin{eqnarray*}
S_0 &=& \frac{1}{K}\sum_{j=0}^{K-1} \w_M^{\xi(j \mu+t)} \\
    &=& \frac{1}{K}\w_M^{\xi t} \sum_{j=0}^{K-1} \w_K^{\xi j} \\
    &=& 0 .
\end{eqnarray*}
Hence when $f$ is constant
the final state of the control register is
$\left| 0^n \right\rangle$,
whereas when $f$ is evenly distributed
the state is orthogonal to $ \left| 0^n \right\rangle $.
Therefore if we discard the auxiliary register and
measure the control register,
then we can determine whether $f$ is nonconstant
or not evenly distributed:
If the outcome of the measurement is $ \left| 0^n \right\rangle $
then $f$ is not evenly distributed and
otherwise $f$ is nonconstant.
As a result the above algorithm can solve
the generalized Deutsch-Jozsa problem by a single evaluation of $f$.

Step (i) transforms the state of the control register
in the equally probable superposition of all input values
using the Walsh-Hadamard operator as usual.
In Step (ii) the information on $f$ is encoded to the phase
of each input value with the help of the auxiliary register
initially prepared in a specific state.
In Step (iii) the phase-encoded information interferes
between input values and the resulting interference pattern
enables the final measurement to give the correct answer.
In the overall procedure we can replace $\cW_n$ by $\QFT$.
In this case the final state becomes
$\frac{1}{N} \sum_{x,y=0}^{N-1}\w_M^{x y} \w_M^{\xi f(x)}
\left| y \right\rangle \otimes  \left| \Psi \right\rangle$
and one can easily check that the same result holds
with this modified algorithm.
We remark that for general positive integers $N$ and $M$
the approximate Fourier transform in \cite{Kitaev} can be used.

Furthermore, when $f$ is evenly distributed
$\mu$ can be found by the quantum period-finding algorithm
which is the core of the quantum factoring algorithm \cite{Shor}.
The application of the quantum Fourier transform to the image of $f$
wipes off the initial shift $t$ and changes its period to
$\frac{M}{\mu}=K$,
so that with high probability we can determine $\mu$
in polynomial time.

Several quantum algorithms have been implemented by NMR quantum
computers
\cite{JM1,Chuang1,Linden,Marx,Collins1,Dorai2,Collins2,Chuang2,%
JM2,JM3,Dorai1,Weinstein,Yannoni}
among which much attention has been paid to the Deutsch-Jozsa
algorithm due to its simplicity whereas the power of a
quantum computer over a classical one can be demonstrated.
In NMR implementation for the Deutsch-Jozsa algorithm
there have been two approaches.
The one \cite{JM1,Chuang1,Linden,Marx}
is the realization of the Cleve's version \cite{Cleve1} that
requires an $n$-qubit control register for storing function arguments
and a one-qubit auxiliary register for function evaluation.
The other one \cite{Dorai2,Collins2} makes use of
the refined Deutsch-Jozsa algorithm in \cite{Collins1}
which is a description of the original Deutsch-Jozsa algorithm
using the conditional phase transform
$\left| x \right\rangle \mapsto (-1)^{f(x)} \left| x \right\rangle$.
The conditional phase transform is a special case of
the {\em $f$-dependent phase transform}
$\cR_{\xi,f}: \left| x \right\rangle \mapsto
\w_M^{\xi f(x)} \left| x \right\rangle$
which plays an important role in most known quantum algorithms
as well as the Deutsch-Jozsa and
the generalized Deutsch-Jozsa algorithms.
The Deutsch-Jozsa and the generalized Deutsch-Jozsa algorithms
are identical except the initial states of the auxiliary registers.
The difference between the initial states of the auxiliary registers
is due to the procedure performing $\cR_{\xi,f}$,
which can be realized by
$\cU_f (\cI \otimes \QFT)
(\left| x \right\rangle \otimes \left| -\xi \right\rangle)$
with the help of the auxiliary register.
Thus if we focus on the control register,
both algorithms are summarized to
$\cW_n \cR_{\xi,f} \cW_n \left| 0^n \right\rangle$.
The $f$-dependent phase transform $\cR_{\xi,f}$
enables us to omit the auxiliary register in the description
of quantum algorithms.
However, in order to implement function-dependent phase transform
without any knowledge on the structure of the given function
we have to evaluate the function on quantum computer.
In this process the auxiliary register is needed
due to the nature of unitary evolution
and all previously known quantum algorithms initialize
the auxiliary register.

We now demonstrate that
a preexisting variant of the original Deutsch-Jozsa algorithm
requires no initialization of the auxiliary one-qubit register,
even though it was previously described with initialization
of the auxiliary register.
We initialize the control register by $ \left| 0^n \right\rangle $.
We let $ \left| \Psi \right\rangle
= a \left| 0 \right\rangle  + b \left| 1 \right\rangle$
be an arbitrary state of the one-qubit auxiliary register
and proceed the following steps:
(i) Apply $\cW_n \otimes \cI$
(ii) Apply $\cU_f$
(iii) Apply $\cI \otimes \sigma_z$
(iv) Apply $\cU_f$
(v) Apply $\cI \otimes \sigma_z$.
Then the state evolves as follows:
\begin{eqnarray}
  \left| 0^n \right\rangle \otimes  \left| \Psi \right\rangle
&\stackrel{\cW_n \otimes \cI}{\longrightarrow}&
 \frac{1}{\sqrt{N}} \sum_{x=0}^{N-1}  \left| x \right\rangle \otimes
 \left| \Psi \right\rangle  \nonumber \\
&\stackrel{\cU_f}{\longrightarrow}&
  \frac{1}{\sqrt{N}} \sum_{x=0}^{N-1}
    \left| x \right\rangle \otimes
    \left( a \left| f(x) \right\rangle
    + b \left| 1+f(x) \right\rangle  \right) \nonumber \\
&\stackrel{\cI \otimes \sigma_z}{\longrightarrow}&
   \frac{1}{\sqrt{N}} \sum_{x=0}^{N-1}
   (-1)^{f(x)}  \left| x \right\rangle \otimes
   \left( a \left| f(x) \right\rangle
   - b \left| 1+f(x) \right\rangle  \right)
   \nonumber \\
&\stackrel{\cU_f}{\longrightarrow}&
  \frac{1}{\sqrt{N}} \sum_{x=0}^{N-1}
  (-1)^{f(x)}  \left| x \right\rangle \otimes
  \left( a \left| 0 \right\rangle
  - b \left| 1 \right\rangle  \right) \nonumber \\
&\stackrel{\cI \otimes \sigma_z}{\longrightarrow}&
    \frac{1}{\sqrt{N}} \sum_{x=0}^{N-1}
     (-1)^{f(x)}  \left| x \right\rangle \otimes
     \left| \Psi \right\rangle .
  \label{Alg:DJ}
\end{eqnarray}
It is noted that this procedure carries out the desired
$f$-dependent phase transform
$\left| x \right\rangle \mapsto (-1)^{f(x)} \left| x \right\rangle$
and recovers the initial state of the auxiliary register.
As in the original Deutsch-Jozsa algorithm
by applying $\cW_n\otimes \cI$ to the final state of (\ref{Alg:DJ})
we can solve the Deutsch-Jozsa problem.
If the auxiliary register is initialized
then the procedure can be simplified.
Starting with the initial state
$ \left| \Psi \right\rangle
= \cW_1  \left| 1 \right\rangle
= \frac{1}{\sqrt{2}}(\left| 0 \right\rangle - \left| 1 \right\rangle)$
we obtain the final state of (\ref{Alg:DJ}) at Step (ii) and
the composite operation of Step (iii), Step (iv), and Step (v)
acts as an identity map.
We note that $a=-b$ is a necessary and
sufficient condition for
$\cU_f \left( \cW_n\otimes \cI \right)
 \left(  \left| 0^n \right\rangle \otimes
 \left| \Psi \right\rangle \right)
= \frac{1}{\sqrt{N}}
\sum_{x=0}^{N-1} (-1)^{f(x)}  \left| x \right\rangle \otimes
\left| \Psi \right\rangle $.

Algorithm (\ref{Alg:DJ}) can be generalized to solve
the generalized Deutsch-Jozsa problem
by using bitwise operations between vectors in $\Z_2^m$
instead of $\Z_{2^m}$.
Let $p:\Z_2^m \longrightarrow \Z_2$ be the parity function and
define a bitwise version of $\cU_f$ as
$\cU_f^\oplus:  \left| x \right\rangle \otimes
\left| y \right\rangle  \mapsto
\left| x \right\rangle \otimes \left| y \oplus f(x) \right\rangle $
where $\oplus$ denotes the bitwise addition in $\Z_2^m$.
We assume that the states of the qubits composing
the auxiliary register are separable.
We initialize the control register by $ \left| 0^n \right\rangle $,
denote the state of the auxiliary register by
$ \left| \Psi \right\rangle = \bigotimes_{j=0}^{m-1}
\left( a_j \left| 0 \right\rangle
+ b_j \left| 1 \right\rangle \right)$,
and proceed the following algorithm:
%
(i) Apply $\cW_n \otimes \cI$
(ii) Apply $\cU_f^\oplus$
(iii) Apply $\cI \otimes \sigma_z^{\otimes m}$
(iv) Apply $\cU_f^\oplus$
(v) Apply $\cI \otimes \sigma_z^{\otimes m}$
(vi) Apply $\cW_n \otimes \cI$.
Then the state evolves as follows:
\begin{eqnarray}
  \left| 0^n \right\rangle \otimes  \left| \Psi \right\rangle
&\stackrel{\cW_n \otimes \cI}{\longrightarrow}&
 \frac{1}{\sqrt{N}} \sum_{x=0}^{N-1}  \left| x \right\rangle
 \otimes \left| \Psi \right\rangle  \nonumber \\
&\stackrel{\cU_f^\oplus}{\longrightarrow}&
  \frac{1}{\sqrt{N}} \sum_{x=0}^{N-1}  \left| x \right\rangle
  \bigotimes_{j=0}^{m-1} \left( a_j \left| f(x)_j \right\rangle
  + b_j \left| 1+f(x)_j \right\rangle  \right)
  \nonumber \\
&\stackrel{\cI \otimes \sigma_z^{\otimes m}}{\longrightarrow}&
   \frac{1}{\sqrt{N}} \sum_{x=0}^{N-1}
   (-1)^{p\circ f(x)}  \left| x \right\rangle
   \bigotimes_{j=0}^{m-1} \left( a_j \left| f(x)_j \right\rangle
   - b_j \left| 1+f(x)_j \right\rangle  \right)
   \nonumber \\
&\stackrel{\cU_f^\oplus}{\longrightarrow}&
  \frac{1}{\sqrt{N}} \sum_{x=0}^{N-1}
  (-1)^{p\circ f(x)}  \left| x \right\rangle
  \bigotimes_{j=0}^{m-1} \left( a_j \left| 0 \right\rangle
  - b_j \left| 1 \right\rangle  \right)
  \nonumber \\
&\stackrel{\cI \otimes \sigma_z^{\otimes m}}{\longrightarrow}&
   \frac{1}{\sqrt{N}} \sum_{x=0}^{N-1}
   (-1)^{p\circ f(x)}  \left| x \right\rangle
   \otimes  \left| \Psi \right\rangle
   \nonumber \\
&\stackrel{\cW_n \otimes \cI}{\longrightarrow}&
 \sum_{y=0}^{N-1}
 \left( \frac{1}{N}\sum_{x=0}^{N-1}
 (-1)^{x \cdot y} (-1)^{p\circ f(x)} \right)
  \left| y \right\rangle  \otimes  \left| \Psi \right\rangle
 \label{Alg:GDJ2}
\end{eqnarray}
where the subscript $j$ represents the $j$-th component of the vector.
Let $S_y'$ be the inner summation in the final state of (\ref{Alg:GDJ2}).
Then when $f$ is constant we have
\begin{eqnarray*}
S_y'
&=& \frac{(-1)^{p\circ f(0)}}{N} \sum_{x=0}^{N-1} (-1)^{x\cdot y}\\
&=& \cases{0 & when $y \ne 0$, \cr (-1)^{p\circ f(0)} & when $y = 0$,}
\end{eqnarray*}
and when $f$ is evenly distributed we get
\begin{eqnarray*}
 S_0'&=& \frac{1}{K}(-1)^{p(t)} \sum_{j=0}^{K-1} (-1)^{p(j \mu)} \\
     &=& \frac{1}{K}(-1)^{p(t)} \sum_{j=0}^{K-1} (-1)^{p(j)} \\
     &=& 0.
\end{eqnarray*}
Here the second equality follows from the fact that
$j \mu \in \Z_2^m$ is an $(m-k)$ left shifts of
a $k$-bit number $j \in \Z_2^k$ with following zeros
where $k=\log_2 K$.
Therefore as before
we can determine whether $f$ is nonconstant or not evenly distributed
and the result still holds
even when $\cW_n$ is replaced by $\QFT$ in Algorithm (\ref{Alg:GDJ2}),
of which the final state is
$\frac{1}{N}\sum_{x,y=0}^{N-1} \w_M^{x y} (-1)^{p\circ f(x)}
 \left| y \right\rangle  \otimes  \left| \Psi \right\rangle $.
As in Algorithm (\ref{Alg:DJ})
the $f$-dependent phase transform
$\left| x \right\rangle \mapsto (-1)^{p\circ f(x)} \left| x \right\rangle$
is obtained at Step (ii) if and only if $a_j=-b_j$ for $j=1,2,\dots,m$.
In this case we can omit Step (iii), Step (iv), and Step (v)
and this simplified algorithm employing the initialization of
the auxiliary register was previously constructed
by Cleve {\em et al.} \cite{Cleve1}
to solve the problem determining
whether $f$ has a constant parity or evenly distributed parities
instead of directly applying the Deutsch-Jozsa algorithm
to the composite function $p \circ f$.

Algorithm (\ref{Alg:GDJ2}) works under the assumption
that the initial state of the auxiliary state is separable,
which is needed in the procedure implementing
function-dependent phase transform.
However, we can eliminate this restriction
by employing the algorithm for function-dependent phase transform
in \cite{CKL} that utilizes two applications of $\cU_f$
as in Algorithm (\ref{Alg:GDJ2}).
In general,
any quantum algorithm that implements function-dependent phase transform
without initializing the auxiliary register
requires at least two evaluations of
the function \cite{CKL}.

This work was supported by the Brain Korea 21 Project.


\end{document}